\pdfoutput=1 

\documentclass[11pt]{article}
\usepackage{amssymb}

\usepackage{amsthm} 

\usepackage{tikz}
\usetikzlibrary{trees}
\usepackage{qtree}

\usetikzlibrary{decorations.pathmorphing}
\usetikzlibrary{decorations.markings}
\usepackage{verbatim}

\usepackage{amsmath}
\usepackage{amsthm} 
\usepackage{amssymb,amsfonts,textcomp}
\usepackage{array}
\usepackage{hhline}
\usepackage{graphicx}

\usepackage[T3,T1]{fontenc} 
\usepackage{txfonts} 

\usepackage{latexsym} 
\usepackage{mathrsfs} 

\usepackage{MnSymbol} 

\usepackage{tikz}
\usetikzlibrary{arrows,shapes,chains,matrix,positioning,scopes}
\usetikzlibrary{decorations.pathmorphing}
\usetikzlibrary{decorations.markings}
\usepgflibrary{plotmarks}
\usetikzlibrary{patterns}

\usepackage{pdfpages}

\usetikzlibrary{mindmap}

\usepackage{graphicx}

\usepackage{multirow}
\usepackage{multicol}

\usepackage{subfig} 

\usepackage{float}
\usepackage{wrapfig}
\restylefloat{figure}
\restylefloat{table}
\usepackage{color, colortbl} 

\usepackage{float}
\restylefloat{figure}

\theoremstyle{definition}

\newcommand{\pot}{o->}

\definecolor{currentcolor}{rgb}{0.8 0.4 0.2}

\tikzstyle{stochasticjumpstyle}=[diamond,draw,fill=white,>=latex,>->,dashed]
\tikzstyle{stochasticPathstyle}=[>=latex,>->,dashed]
\tikzstyle{stochasticNodestyle}=[ellipse,inner sep=1pt,text=.,fill=.!20]
\tikzstyle{blankstyle}=[fill=white,inner sep=1pt]

\def\SnakeSegLen{0.6em}
\def\SnakeAmp{0.11em}
\def\PrePostLen{5mm}

\tikzstyle{sendstyle}=[dashed,line width=1.1pt]

\tikzstyle{splitstyle}=[circle,draw]

\tikzstyle{receivestyle}=[>->,line width=1.1pt,decorate, decoration={zigzag,segment length=\SnakeSegLen, amplitude=\SnakeAmp, pre length=\PrePostLen, post=curveto, post length=\PrePostLen},text=black]

\tikzstyle{receivesigstyle}=[draw,inner sep=2pt,fill=pink!20]
\tikzstyle{receivesigstyle3}=[draw,inner sep=2pt, fill=white]
\tikzstyle{receivesigstyle2}=[ellipse,shade, draw,double,fill=red!10]

\tikzstyle{sendsigstyle}=[diamond,draw,inner sep=1pt, text=black, fill=yellow!80]
\tikzstyle{sendsigstyle3}=[circle,draw, ball color=white]
\tikzstyle{sendsigstyle2}=[diamond,draw,double, inner sep=1pt, fill=white]

\tikzstyle{snakesendstyle}=[*->, decorate, decoration={snake, segment length=\SnakeSegLen, amplitude=\SnakeAmp,  pre length=\PrePostLen, post=curveto, post length=\PrePostLen}]

\tikzstyle{snakesendstyle1}=[line width=1.1pt, decorate, decoration={snake,segment length=\SnakeSegLen, amplitude=\SnakeAmp}]

\tikzstyle{snakesendstyle3}=[decorate, decoration={markings, mark=at position .75 with {\arrow[red,line width=5mm]{>}}, snake, segment length=\SnakeSegLen, amplitude=\SnakeAmp,  pre length=\PrePostLen, post=curveto, post length=\PrePostLen}]

\tikzstyle{snakesendstyle2}=[decorate, decoration={ zigzag,segment length=\SnakeSegLen, amplitude=\SnakeAmp, line around/.style={decoration={pre length=\PrePostLen,post length=\PrePostLen}}}]

\newcounter{foo}
\usepackage{hyperref}

\makeatletter
\AtBeginDocument{\let\nameref\HyPsd@nameref}

\makeatother

\colorlet{anglecolor}{green!50!black}

\definecolor{darkgreen}{rgb}{0 0.6  0}

\definecolor{turquoise}{rgb}{0 0.41 0.41}
\definecolor{rouge}{rgb}{0.79 0.0 0.1}
\definecolor{vert}{rgb}{0.15 0.4 0.1}
\definecolor{mauve}{rgb}{0.6 0.4 0.8}
\definecolor{violet}{rgb}{0.58 0. 0.41}
\definecolor{orange}{rgb}{0.8 0.4 0.2}
\definecolor{bleu}{rgb}{0.39, 0.58, 0.93}

\definecolor{darkross}{rgb}{0.008,0.412,0.471}
\definecolor{middleross}{rgb}{0.012,0.580,0.663}
\definecolor{lightross}{rgb}{0.016,0.749,0.855}
\definecolor{darkblue}{rgb}{0.067,0.008,0.471}
\definecolor{middleblue}{rgb}{0.094,0.012,0.663}
\definecolor{lightblue}{rgb}{0.122,0.016,0.855}
\definecolor{darkpurple}{rgb}{0.471,0.008,0.412}
\definecolor{middlepurple}{rgb}{0.663,0.012,0.580}
\definecolor{lightpurple}{rgb}{0.855,0.016,0.749}
\definecolor{darkbrown}{rgb}{0.471,0.067,0.008}
\definecolor{middlebrown}{rgb}{0.663,0.094,0.012}
\definecolor{lightbrown}{rgb}{0.855,0.122,0.016}
\definecolor{darkolive}{rgb}{0.412,0.471,0.008}
\definecolor{middleolive}{rgb}{0.580,0.663,0.012}
\definecolor{lightolive}{rgb}{0.749,0.855,0.016}
\definecolor{darkgreen}{rgb}{0.008,0.417,0.067}
\definecolor{middlegreen}{rgb}{0.012,0.663,0.094}
\definecolor{lightgreen}{rgb}{0.016,0.855,0.122}
\definecolor{darkocre}{rgb}{0.471,0.298,0.008}
\definecolor{middleocre}{rgb}{0.663,0.420,0.012}
\definecolor{lightocre}{rgb}{0.855,0.541,0.016}

    \definecolor{lightblue}{rgb}{0,0,.7}
    \definecolor{orange}{rgb}{1,.7,0}
    \definecolor{darkorange}{rgb}{1,.4,0}
    \definecolor{darkgreen}{rgb}{0,.5,0}
    \definecolor{darkblue}{rgb}{0,0,.4}
    \definecolor{darkred}{rgb}{.4,0,0}
    \definecolor{gray}{rgb}{.2,.2,.2}
    \definecolor{darkgray}{rgb}{.2,.2,.2}
    \definecolor{shadecolor}{gray}{0.925}
    
\definecolor{darkred}{rgb}{0.65,0,0}
\definecolor{darkblue}{rgb}{0,0,.65}
\definecolor{darkgreen}{rgb}{0,0.5,0}
\definecolor{orange}{rgb}{1,.75,.25}
\definecolor{aqua}{rgb}{0,.25,.75}
\definecolor{grey}{rgb}{.5,.5,.5}
\definecolor{brown}{rgb}{.51,.35,.18}

\definecolor{lightblue}{rgb}{.3,.5,1}
\definecolor{orange}{rgb}{1,.7,0}
\definecolor{darkorange}{rgb}{1,.4,0}
\definecolor{darkgreen}{rgb}{0,.4,0}
\definecolor{darkblue}{rgb}{0,0,.4}
\definecolor{darkred}{rgb}{.56,0,0}
\definecolor{gray}{rgb}{.3,.3,.3}
\definecolor{darkgray}{rgb}{.2,.2,.2}

\definecolor{blue}{rgb}{0,0,1}
\definecolor{red}{rgb}{1,0,0}
\definecolor{pink}{rgb}{.933,0,.933}
\definecolor{green}{rgb}{0.133,0.545,0.133}
\definecolor{shadecolor}{gray}{0.925}

\definecolor{DarkBlue}{rgb}{0.000,0.000,0.545}
\definecolor{DarkChocolate}{rgb}{0.400,0.200,0.000}
\definecolor{DarkCyan}{rgb}{0.000,0.545,0.545}
\definecolor{DarkGoldenrod}{rgb}{0.720,0.525,0.044}
\definecolor{DarkGray}{rgb}{0.664,0.664,0.664}
\definecolor{DarkGreen}{rgb}{0.000,0.392,0.000}
\definecolor{DarkGrey}{rgb}{0.664,0.664,0.664}
\definecolor{DarkKhaki}{rgb}{0.740,0.716,0.420}
\definecolor{DarkLavender}{rgb}{0.400,0.200,0.600}
\definecolor{DarkMagenta}{rgb}{0.545,0.000,0.545}
\definecolor{DarkOliveGreen}{rgb}{0.332,0.420,0.185}
\definecolor{DarkOrange}{rgb}{1.000,0.550,0.000}
\definecolor{DarkOrchid}{rgb}{0.600,0.196,0.800}
\definecolor{DarkPeriwinkle}{rgb}{0.400,0.400,1.000}
\definecolor{DarkPurpleBlue}{rgb}{0.400,0.000,0.800}
\definecolor{DarkRed}{rgb}{0.545,0.000,0.000}
\definecolor{DarkRoyalBlue}{rgb}{0.000,0.200,0.800}
\definecolor{DarkSalmon}{rgb}{0.912,0.590,0.480}
\definecolor{DarkSeaGreen}{rgb}{0.560,0.736,0.560}
\definecolor{DarkSlateBlue}{rgb}{0.284,0.240,0.545}
\definecolor{DarkSlateGray}{rgb}{0.185,0.310,0.310}
\definecolor{DarkSlateGrey}{rgb}{0.185,0.310,0.310}
\definecolor{DarkSmoke}{rgb}{0.920,0.920,0.920}
\definecolor{DarkTurquoise}{rgb}{0.000,0.808,0.820}
\definecolor{DarkViolet}{rgb}{0.580,0.000,0.828}
\definecolor{DeepPink}{rgb}{1.000,0.080,0.576}
\definecolor{DeepSkyBlue}{rgb}{0.000,0.750,1.000}

\tikzstyle{mystyle}=[scale= \PicSize,  
baseline=(current bounding box.center),
nodedescr/.style={fill=white,inner sep=2.5pt},
nodedescr2/.style={draw,circle,fill=white},
description/.style={fill=white,inner sep=2pt},
cancer loop/.style={preaction={draw=white, -,line width=6pt}, line width=1.1pt},
pot1 loop/.style={preaction={draw=white, -,line width=6pt}, red, line width=1.1pt},
pot2 loop/.style={preaction={draw=white, -,line width=6pt}, green, line width=1.1pt},
pot1/.style={preaction={draw=white, -,line width=6pt}, gray, solid, line width=1pt}, 
pot2/.style={preaction={draw=white, -,line width=6pt}, gray, solid, line width=1pt},
inPot/.style={ out=45,in=90,distance=3cm, min distance=2cm, line width=1.1pt, black!75},
inPot1/.style={ out=45,in=120,distance=3cm, min distance=2cm, line width=1.1pt, black!75},
inPot2/.style={ out=-45,in=-120,distance=3cm, min distance=2cm, line width=1.1pt, black!75},
in100Pot1/.style={ out=80,in=90,distance=3cm, min distance=2cm, line width=1.1pt, darkgreen},
in100Pot2/.style={ out=-80,in=-90,distance=3cm, min distance=2cm, line width=1.1pt, brown},
in2Pot1/.style={ out=45,in=90,distance=3cm, min distance=2cm, line width=1.1pt, black!75},
in2Pot2/.style={ out=-65,in=-90,distance=3cm, min distance=2cm, line width=1.1pt, black!75},
selfloop1/.style={ out=45,in=120,distance=6cm, min distance=4cm, line width=1.1pt, red},
selfloop2/.style={ out=-45,in=-120,distance=6cm, min distance=4cm, line width=1.1pt, blue},
loop1red/.style={ out=45,in=120,distance=6cm, min distance=4cm, line width=1.1pt, red},
loop2red/.style={ out=-45,in=-80,distance=6cm, min distance=4cm, line width=1.1pt, red},
cross line/.style={preaction={draw=white, -,	line width=6pt}},
jump1/.style={ out=45,in=135,distance=4cm, min distance=2cm, line width=1.1pt, red, dashed},
jump2/.style={ out=-45,in=-135,distance=4cm, min distance=2cm, line width=1.1pt, blue, dashed},
jumpover/.style={ out=80,in=90,distance=4cm, min distance=3cm, line width=1.1pt, black, dashed},
stochastic jump/.style={>=latex,>->,dashed, line width=1.1pt, green}
]

\def\PicSize{ 0.5} 

\numberwithin{equation}{section}

\usepackage[margin=1.4in]{geometry} 

\usepackage[parfill]{parskip}    

\begin{document}

\title{Stem Cells: \\The Good, the Bad and the Ugly}


\author{Eric Werner \thanks{Balliol Graduate Centre, Oxford Advanced Research Foundation (http://oarf.org).  We gratefully acknowledge the use of Cellnomica's CancerCAD software to construct the cancer and stem cell networks used to model and simulate all the multicellular processes that generated the {\em in silico} cancers described and illustrated in this paper.  \copyright Werner 2016.  All rightsreserved. }\\
University of Oxford\\
Department of Physiology, Anatomy and Genetics, \\
and Department of Computer Science, \\
Le Gros Clark Building, 
South Parks Road, 
Oxford OX1 3QX  \\
email:  eric.werner@dpag.ox.ac.uk\\
}

\date{ } 

\maketitle

\thispagestyle{empty}

\begin{center}
\textbf{Abstract}

\begin{quote}
\it
Cancer stem cells are controlled by developmental networks that are often topologically indistinguishable from normal, healthy stem cells.  The question is why cancer stem cells can be both phenotypically distinct and have morphological effects so different from normal stem cells.  The difference between cancer stem cells and normal stem cells lies not in differences their network architecture, but rather in the spatial-temporal locality of their activation in the genome and the resulting expression in the body.  The metastatic potential cancer stem cells is not based primarily on their network divergence from normal stem cells, but on non-network based genetic changes that enable the evolution of gene-based phenotypic properties of the cell that permit its escape and travel to other parts of the body.  Stem cell network theory allows the precise prediction of stem cell behavioral dynamics and a mathematical description of stem cell proliferation for both normal and cancer stem cells.  It indicates that the best therapeutic approach is to tackle the highest order stem cells first, otherwise spontaneous remission of so called cured cancers will always be a danger.  Stem cell networks point to a pathway to new methods to diagnose and cure not only stem cell cancers but cancers generally. 
\end{quote}
\end{center}
{\bf Key words}: {\sf  stem cells, stem cell networks, cene, cenome, developmental control networks, cancer stem cells, stochastic stem cell networks, stochastic cancer stem cell networks, metastatic hierarchy, linear networks, geometric cancer networks, systems biology, computational biology, multiagent systems, muticellular modeling, simulation, cancer modeling, cancer simulation, CRISPR, genome editing, cancer cure}

\pagebreak

\pagenumbering{roman}
\setcounter{page}{1}
\tableofcontents
\newpage
\pagenumbering{arabic}

\section{Introduction}

How do healthy and cancer stem cells differ?
In my stem cell networks paper \cite{Werner2016b} there is considerable overlap with my cancer networks paper \cite{Werner2011b}. The reason is that cancer stem cell networks and healthy stem cell networks are in many ways topologically indistinguishable. Yet, the phenotype of a cancer stem cell and the cells it generates may differ greatly from healthy stem cells even if they are governed by equivalent network types.  

One reason for the difference is if the spatial-temporal activation or locality of cancer stem cell networks within the global network that controls the development of the organism is abnormal.  If a stem cell network is inappropriately expressed in the global developmental control network it can result in very different cell phenotype as well as an abnormal locus of tumor growth.  So too, if a stem cell network is inappropriately activated in any cell, then abnormal tumor like growth may result in that location of the body. 

\section{Cancer stem cell formation}
A linear, 1st-order stem cell can form anywhere in a given developmental control network when a single loop is formed that generates two daughter cells, one that whose downstream offspring loops back to the original stem cell and one that goes down another pathway that does not loop back. The phenotype of the resulting tumor depends on the location of the initiation point of the loop, the loop pathway and the pathway of the non-looping terminal network.  For example, $A$ divides into two daughter cells $B$ and $C$ where $B$ is controlled by network $N_B$ and $C$ is controlled by network $N_C$ and where $N_B$ loops back to $A$ eventually. And $N_C$ is a terminal developmental network. The simplest case is where the networks consist of just one node, e.g.,  $N_B = A$ and $N_C = T$ where $T$ is some terminal cell state. The cell at node $A$ then just generates terminal cells of type $T$ and loops back to itself.  But if $N_B$ and/or $N_C$ are more complex networks then one can get very odd stem cell cancers.  If then a second mutational transformation results in yet a second loop back to $A$ in $N_C$ the second daughter cell's network, we suddenly get exponential, explosive cell proliferation (assuming no conditional activation such as dependence on cell-cell signaling).  

\section{Activation potential}
Activation of a hidden stem cell network is necessary for stem cell development or cancer development to initiate.  Once initiated it continues {\em ad infinitum} unless held in check by some necessary condition such as a cell signal or other source. Thus there may exist many passive stem cell and exponential networks in the genome but that are inactive and not accessible in the normal execution of the global developmental control network.  Environmental influences can lead to the activation of such passive networks resulting in tumors whose properties depend on the location and topology of the network.  Further mutations in such an activated network can have severe consequences by changing a linear stem cell network into an exponential network \cite{Werner2016b, Werner2011b} as illustrated in Figure \ref{fig:LinToExpoTransform} below.

\section{A potential danger with all stem cells}
\label{sec:StemToExpo}
Thus, even healthy normal stem cells carry with them an inherent danger. The reason is that mutations that transform developmental networks can convert a normal, healthy stem cell network into a dangerous exponential network by the mere addition of second backwards loop.  The result can be catastrophic with explosive tumor growth.  This actually happens in some brain cancers (glioblastoma) and bone marrow cancers (acute myeloid leukemia AML and acute lymphoblastic leukemia) where slow growing normal stem cells responsible for the regenerating healthy cells is suddenly transformed and become explosive (see  \cite{Mukherjee2010} for an informal account). 

\begin{figure}[H]
\subfloat[We start with 2 stem cells]{
\includegraphics[scale=0.55]{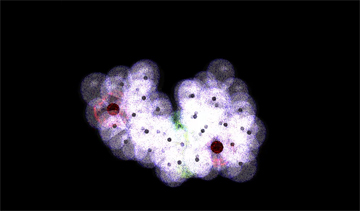}
\label{fig:LinToExpoLin}
}
\hspace{0.3cm}
\subfloat[One extra loop mutation leads to explosive growth]{
\includegraphics[scale=0.49]{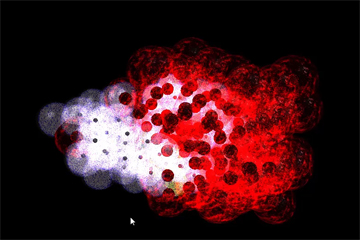}
\label{fig:LinToExpo}
}
\caption{
{\bf A stem cells gone bad.} A transformative mutation that leads to a upstream loop in the right stem cell transforms the stem cell network into an exponential network \cite{Werner2016b, Werner2011b}. This results in explosive growth as seen in bone marrow cancers (acute myeloid leukemia AML and acute lymphoblastic leukemia) and brain cancers (glioblastoma).   
}
\label{fig:LinToExpoTransform}
\end{figure}

\section{Inadequacy of the standard gene-centered paradigm of how stem cells work}
The gene-centered model of cancer which dominates current scientific thinking has difficulty explaining this phenomenon if at all because it views cancer as uncontrolled growth. This paradigm gives no insight into how to control cancer except by finding ways to destroy cancer cells. The therapy consists of cutting out, killing by chemicals or radiation, gene-centered therapies and/or using the bodies own defenses (immunotherapy).  Immunotherapy is the latest hope and it appears to work for some cancers.  While we all hope it will work for all cancers, it is still just a hope. Certainly, we have made enormous strides in treating and managing cancer. 

However, in spite of its successes, the gene-centered paradigm and the resulting research has inherent limits in furthering our understanding of the underlying mechanisms in cancer cells. It has no theoretical terminology to even describe the problem.  Hence, the gene-centered paradigm actually hinders finding methods that will lead to a real cure for all cancers (see section \ref{sec:PathToCure}).

A clear example in Mukherjee's book {\em Cancer the Emperor of All Maladies}. In it he writes:
\begin{quote}
In acute lymphoblastic leukemia, as in some other cancers, the overproduction of cancer cells is combined with a mysterious arrest in the normal maturation of cells. Lymphoid cells are thus produced in vast excess, but, unable to mature \ldots \cite{Mukherjee2010}. 
\end{quote}

In the Stem Cell Networks paradigm it is just a consequence of the structure of the network itself, which loops back to the generating cell's state before the cell can mature into a terminal cell.  The gene-centered view has no theoretical concept to even formulate  this phenomenon let alone to explain it\footnote{The authors Hanahan and Weinberg who wrote the influential review of the hallmarks of cancer \cite{Hanahan2000} recently updated their account to include gene regulatory networks (GRNs) \cite{Hanahan2011}.  However, as is shown in \cite{Werner2013, Werner2015} there are inherent combinatorial limits to GRNs making them unlikely to give a realistic explanation of cancer networks of the type described in \cite{Werner2011b, Werner2016b}.}

\section{The network paradigm of stem cells}
In contrast my network theory of stem cells and more generally cancer networks gives an exact explanation of these phenomena when stem cells go bad. The stem cell network paradigm gives detailed models that can be mathematically formalized in a variety of ways, including differential equations  \cite{McDuffie2014, Manley2014, Epelle2013, Varela2015, Bodine2016}.  The networks have been used to model tumor therapies including radiation, proton therapy.  While these models use the simplest possible stem cell network model as a starting point, in practice they use the equivalent of a stochastic stem cell network with stochastic dedifferentiation such as the following: 

\label{sec:G1SD}
\begin{figure}[H]
\begin{tikzpicture}[style=mystyle]
\matrix (m) [matrix of math nodes, row sep=3em,
column sep=3em, text height=1.5ex, text depth=0.25ex]
{ \vphantom{a} & \vphantom{b}  &  \vphantom{c}  & \vphantom{c}  & \\
 A  &&  B  &&  T  &&  \\
  \vphantom{a} &   \vphantom{j} & \vphantom{b} &  \vphantom{c}  & \\ };
 \path[\pot]
(m-2-1) edge [selfloop1, cross line] node[nodedescr] {$ a $} (m-2-1)
(m-2-1) edge [inPot2, in=-110, blue, cross line] node[nodedescr] {$j$} (m-2-3); 
\path[stochasticPathstyle]
(m-2-3) edge [inPot2, out=-80, in=-90,distance=4cm, green, cross line] node[stochasticNodestyle] {$p$} (m-2-1) 
(m-2-3) edge [in2Pot1,in=90,out=80,distance=4cm,green, cross line] node[stochasticNodestyle] {$1-p$} (m-2-5); 
\path[solid, line width=6pt]
(m-2-1) edge (m-2-3)
(m-2-3) edge (m-2-5);
\end{tikzpicture}
\caption{
    {\bf Network G1SD: A 1st-order geometric cancer with stochastic dedifferentiation.}  This example is adapted from \cite{Werner2016b}.
  }
    \label{fig:G1SD}
\end{figure}
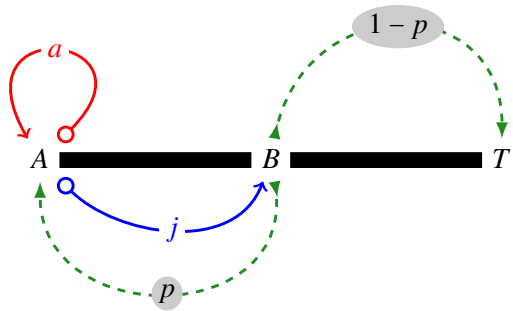 

In this linear network, a cell of type $A$ is a 1st-order, linear stem cell. It produces progenitor cells of type $B$ that have a stochastic dedifferentiation potential.  The cell $B$ can either dedifferentiate to its parent state $A$ or differentiate into the terminal cell $T$.  The lower the probability $p$ of dedifferentiation, the greater the number of terminal cells in the cell population generated by the stem cell $A$. The higher the probability $p$ the more the network approaches the topology of a pure exponential network with explosive cell proliferation.  Thus the topology of the network is a pragmatic topology (see below) that depends on the probability distribution.  The effect of the probability distribution is shown in the following example adapted from \cite{Werner2016b}:

\begin{figure}[H]
\subfloat[Probability $p$ = 0.52, gives 4\% stem cells]{
\includegraphics[scale=0.4]{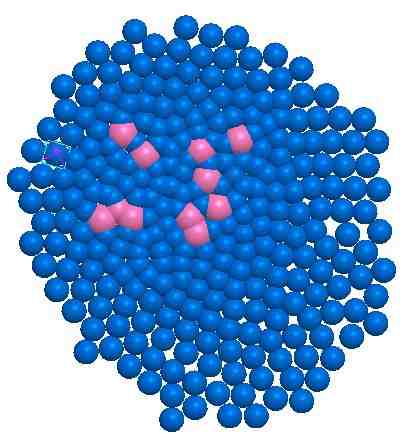}
\label{fig:StocStemSubfig1}
}
\hspace{0.3cm}
\subfloat[When $p$ = 0.7 gives 36\% stem cells]{
\includegraphics[scale=0.4]{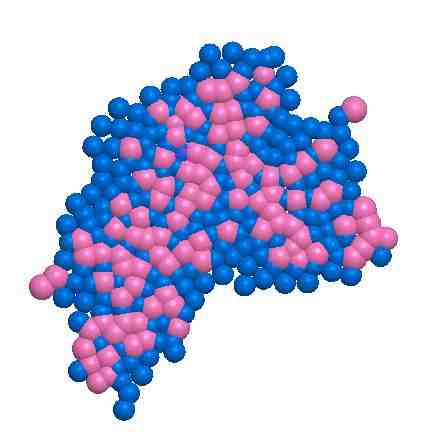}
\label{fig:StocStemSubfig2}
}
\hspace{0.3cm}
\subfloat[Probability $p$ = 0.9 gives 80\% stem cells]{
\includegraphics[scale=0.4]{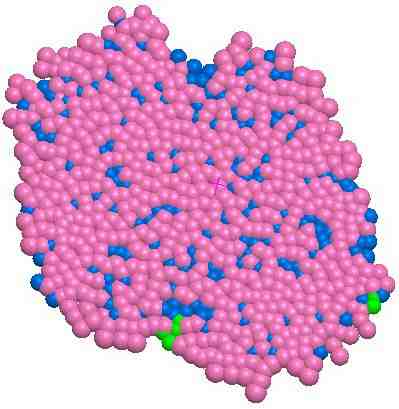}
\label{fig:StocStemSubfig3}
}
\caption{
{\bf Three views of stem cells under different probabilities.} All three tumors are  controlled by a similar network as in (\autoref{fig:G1SD} ).  It shows the growth response to changes in the probability $p$.   As the probability $p$ is increased, more and more stem cells are generated due to increasing dedifferentiation of the daughter cells to their parent stem cell.  The stochastic exponential network begins to dominate as $p$ approaches 1. Example is adapted from the classic Till network \cite{Till1964} as formalized in \cite{Werner2016b}.  
}
\label{fig:StocStemProb}
\end{figure}

\subsection{Stochastic linear networks formalized as differential equations}
The stochastic dedifferentiation is modeled by an extra term in the differential equation that allows a second loop from the terminal cell to occur occasionally. Where 'occasionally' is dependent on the magnitude of the extra term.  The stochastic dedifferentiation allows the modeling of tumors that exhibit resistance to therapy or that spontaneously reoccur after some time of remission.  The therapeutic success is dependent on the network that is actually being modeled \cite{McDuffie2014, Manley2014, Epelle2013, Varela2015, Bodine2016}. 

\subsection{Pragmatic-stochastic topology of networks}
The important point is that the probability distribution over a stochastic stem cell network determines the pragmatic topology of the network.  By {\em pragmatic topology} I mean the range of network architectures between the ideal network as defined by its possible paths (the phase space of the network) versus all the possible executions of the network given the probability distribution. Thus we have a range of possible outcomes from a truly deterministic, linear stem cell network (where the dedifferentiation probabilities are $0$) all the way to a deterministic exponential network (where the dedifferentiation probabilities are $1$). 

\subsection{When non-network genes are important}
There are further differences between healthy and stem cell networks that have to do with non-networks mutations as happens, for example, when genetic mutations in genes that enable cell migration and  metastases to form.  

\section{Which cancer stem cells are the most dangerous?}
Beyond the ever present dangerous potential of cancer stem cell networks to transform into exponential networks (see section  \ref{sec:StemToExpo}), the order of the stem cell network determines how dangerous the metastatic stem cell is. A first order stem cell (controlled by a 1st order, linear stem cell network) will only produce non-proliferative, and in this sense, harmless, terminal cells\footnote{A complication is if there is stochastic dedifferentiation in presumptive terminal cells then the probabilities determine the effective topology of the stem cell network leading to degrees of possible non-linear growth.}.

A second-order stem cell network will produce 1st order stem cells which if they metastasize will produce unwanted cyst-like growths. 

More severe is the effect of metastatic 3rd-order stem cell networks which produce 2nd-order stem cells. Then the metastatic 2nd order cells will also produce 1st-oder stem cells leading to faster tumor growth and possibly more metastases. 

With this network understanding of higher orders of stem cells, it is clear that to treat such cancers that the most important cancer cell to kill first is the stem cell of the highest order. 

\section{Precise predictions of dynamics of proliferation}
The stem cell network theory also gives precise predictions of the upper bounds of cell proliferation in real tumors and cancers.  As I showed in the stem cell networks paper the dynamics and growth in an ideal context of discrete space time is related to the form classic Greek figurative numbers and Platonic objects, as well as Pascal's Triangle and the coefficients of the Binomial Theorem.  This gives the upper bounds of stem cell growth in continuous space-time in real tumors with real cells. 

\section{Stem cell networks point to new methods to cure cancers}
\label{sec:PathToCure}
Unlike the gene-centered paradigm which views cancer as uncontrolled growth, stem cell networks are founded on the view that cancer like embryonic development is a highly controlled process \cite{Varela2015, Werner2011b, Werner2016b}.  Given that stem cell networks are encoded in the genome, then a path to cure cancer becomes possible.  

\begin{enumerate}
\item First, we need to crack the cancer code \cite{Werner2003b, Werner2005}. This means we need to find the translation that interrelates encoding of {\em in silico} stem cell networks with natural stem cell networks encoded in the DNA of genomes. Once, that is achieved, we have many options by which stem cell cancers can be cured.  

\item We start by searching the genome of cancer cells to find the cancer network that controls the cancer cell's  proliferation. Once, the controlling cancer network is found there are a number of possible options:  

\begin{enumerate}
\item New methods can be developed to kill all cells with the active cancer network. A cautionary note: Killing cells with cancer networks has to be done with care, since some cancer networks, such as networks involved in bilateral breast cancer are most likely in the genome at birth and hence reside in every cell of the body. Killing all cells that contain the cancer network would kill every cell in the body!  

\item Methods can be developed to edit the cancer network so as to inactivate it thereby stopping cell proliferation\footnote{The existence of the CRISPR biotechnology to edit genomes shows that it is in principle possible to edit cancer networks \cite{Doudna2014, Doudna2013a, Doudna2013b}. This method combines search with editing. }.  

\item If the cancer network is a mutated normal stem cell network, methods can be developed to edit the cancer network transforming back into its previously healthy state. This method could be used to stop explosive cancers governed by stem cells whose linear stem cell network has been transformed into an exponential network.  
\end{enumerate}
\end{enumerate}

As the work on CRISPR by Doudna and her team indicates \cite{Doudna2014, Doudna2013a, Doudna2013b}, future bioengineers will, of course, find and invent many new methods to transform such cancer networks or to carefully target and kill cells with such networks.  In that sense, it is a good thing that cancer stem cell networks are so similar to normal, healthy stem cell networks since it gives  the understanding that is needed to precisely model and simulate such cancers. And it offers a roadmap that may ultimately lead to a cure for such cancers. 

\pagebreak
 \nocite{Mukherjee2010}
 \nocite{McDuffie2014, Manley2014, Epelle2013, Varela2015, Suen2006}
 \nocite{Werner2003b, Werner2005, Werner2007a, Werner2009,Werner2010,Werner2011b,Werner2016b}
\nocite{Tyson1991, Tyson2003}
\nocite{Dick2010, Dick2009, Dick2008}
\nocite{Shackleton2010, Shackleton2010a, Shackleton2010b, Shackleton2010c}
\nocite{Sonnenschein1999}
\nocite{Bizzarri2008}

\addcontentsline{toc}{section}{References}
\begin{multicols}{2}

\footnotesize 
\bibliographystyle{abbrv}
\bibliography{StemCells} 
\end{multicols}
\end{document}